\documentclass[amsfonts,amsmath,amssymb,prl,twocolumn,superscriptaddress]{revtex4-2}

\usepackage[compact]{titlesec}
\usepackage[utf8]{inputenc}
\usepackage{graphicx}
\usepackage{dcolumn}
\usepackage{bm}
\usepackage{natbib}
\usepackage{float}
\usepackage{amsmath}
\usepackage[dvipsnames]{xcolor}
\usepackage{midfloat}
\usepackage{datetime2}
\usepackage{ragged2e}
\usepackage{lineno}
\usepackage{indentfirst}
\usepackage{gensymb}
\usepackage{comment}

\makeatletter

\def\maketitle{
    \@author
    \title@column
    \titleblock@produce
    \suppressfloats[t]
}

\def\makesuptitle{
    \centering{\textbf{\large{Supplementary Material for\\}}}
    \@author
    \titleblock@produce
    \suppressfloats[t]
}

\makeatother

\newcommand{\beginsupplement}{
    \clearpage
    \onecolumngrid
    \makesuptitle
    \setcounter{table}{0}
    \renewcommand{\thetable}{S\arabic{table}}
    \setcounter{figure}{0}
    \renewcommand{\thefigure}{S\arabic{figure}}
    \setcounter{equation}{0}
    \renewcommand{\theequation}{S\arabic{equation}}
    \setcounter{secnumdepth}{0}
}

\begin{document}

\title{Quantitative determination of twist angle and strain \\in Van der Waals moir\'e superlattices}

\author{Steven J. Tran}
\affiliation{Stanford Institute for Materials and Energy Sciences, SLAC National Accelerator Laboratory, Menlo Park, CA 94025}
\affiliation{Department of Physics, Stanford University, Stanford, CA 94305}

\author{Jan-Lucas Uslu}
\affiliation{JARA-FIT and 2nd Institute of Physics, RWTH Aachen University, 52074 Aachen, Germany, EU}

\author{Mihir Pendharkar}
\affiliation{Stanford Institute for Materials and Energy Sciences, SLAC National Accelerator Laboratory, Menlo Park, CA 94025}
\affiliation{Department of Materials Science and Engineering, Stanford University, Stanford, CA 94305}

\author{Joe Finney}
\affiliation{Stanford Institute for Materials and Energy Sciences, SLAC National Accelerator Laboratory, Menlo Park, CA 94025}
\affiliation{Department of Physics, Stanford University, Stanford, CA 94305}

\author{Aaron L. Sharpe}
\affiliation{Stanford Institute for Materials and Energy Sciences, SLAC National Accelerator Laboratory, Menlo Park, CA 94025}
\affiliation{Department of Physics, Stanford University, Stanford, CA 94305}

\author{\\Marisa Hocking}
\affiliation{Stanford Institute for Materials and Energy Sciences, SLAC National Accelerator Laboratory, Menlo Park, CA 94025}
\affiliation{Department of Materials Science and Engineering, Stanford University, Stanford, CA 94305}

\author{Nathan J. Bittner}
\affiliation{Independent Researcher}

\author{Kenji Watanabe}
\affiliation{Research Center for Electronic and Optical Materials, National Institute for Materials Science, 1-1 Namiki, Tsukuba 305-0044, Japan}

\author{Takashi Taniguchi}
\affiliation{Research Center for Materials Nanoarchitectonics, National Institute for Materials Science,  1-1 Namiki, Tsukuba 305-0044, Japan}

\author{\\Marc A. Kastner}
\affiliation{Stanford Institute for Materials and Energy Sciences, SLAC National Accelerator Laboratory, Menlo Park, CA 94025}
\affiliation{Department of Physics, Stanford University, Stanford, CA 94305}
\affiliation{Department of Physics, Massachusetts Institute of Technology, Cambridge, MA 02139}

\author{Andrew J. Mannix}
\affiliation{Stanford Institute for Materials and Energy Sciences, SLAC National Accelerator Laboratory, Menlo Park, CA 94025}
\affiliation{Department of Materials Science and Engineering, Stanford University, Stanford, CA 94305}

\author{David Goldhaber-Gordon}
\email{goldhaber-gordon [at] stanford [dot] edu}
\affiliation{Stanford Institute for Materials and Energy Sciences, SLAC National Accelerator Laboratory, Menlo Park, CA 94025}
\affiliation{Department of Physics, Stanford University, Stanford, CA 94305}

\date{12 June 2024}

\begin{abstract}
Scanning probe techniques are popular, non-destructive ways to visualize the real space structure of Van der Waals moir\'es. The high lateral spatial resolution provided by these techniques enables extracting the moir\'e lattice vectors from a scanning probe image. We have found that the extracted values, while precise, are not necessarily {\em accurate}. Scan-to-scan variations in the behavior of the piezos which drive the scanning probe, and thermally-driven slow relative drift between probe and sample, produce systematic errors in the extraction of lattice vectors. In this Letter, we identify the errors and provide a protocol to correct for them. Applying this protocol to an ensemble of ten successive scans of near-magic-angle twisted bilayer graphene, we are able to reduce our errors in extracting lattice vectors to less than 1\%. This translates to extracting twist angles with a statistical uncertainty less than $0.001$° and uniaxial heterostrain with uncertainty on the order of $0.002$\%. 
\end{abstract}

\maketitle
The electronic properties of Van der Waals (VdW) moir\'e heterostructures depend sensitively on the relative twist between layers of the stack. In experimental samples, the final twist angle is often different than intended and/or spatially nonuniform. Testing approaches to improve this situation will require a rapid way to assess structure during or after stacking. Several scanning probe techniques implemented on commercial atomic force microscope (AFM) platforms \cite{pendharkar_torsional_2024, lee_ultrahigh-resolution_2020, kapfer_programming_2023, huang_imaging_2021, mcgilly_visualization_2020, zhang_dynamic_2023} can image nanometer-scale moir\'e superlattices in real space. To extract quantitative structural information from such scanning probe imaging, distortions in the raw spatial maps due to scanning probe hardware and thermal drift \cite{yothers_real-space_2017, rahe_vertical_2010, salmons_correction_2010} must be accounted for and removed.

In this Letter, we describe how to extract quantitative information about local moir\'e superlattice structure from scanning probe images. We use images of twisted bilayer graphene (TBG) acquired through torsional force microscopy (TFM), but the method we present should work for any moir\'e superlattice imaged by any scanning probe technique. We first show the importance of using a coordinate system which reflects the actual rather than intended motion of the scanning probe when scanning with position feedback disabled. We then present a protocol to correct image distortions caused by the slow relative drift between sample and probe due to thermal expansion. Layers in moir\'e heterostructures generally have some heterostrain, i.e. each layer is strained relative to its neighbor. Scanning tunneling microscopy measurements have shown this heterostrain to have typical magnitude of a few tenths of a percent \cite{mesple_heterostrain_2021, kerelsky_maximized_2019, liu_spectroscopy_2021, benschop_measuring_2021}. Heterostrain on this scale can dramatically influence the electronic properties of moir\'e superlattices \cite{bi_designing_2019, parker_strain-induced_2021, gao_heterostrain-enabled_2021}. If we assume that heterostrain is purely uniaxial, our analysis of images allows us to extract both the TBG twist angle and the uniaxial heterostrain with their respective uncertainties. Additional information could allow distinguishing biaxial heterostrain from a shift in twist angle. 

We target twist angle uncertainty of 0.01° and heterostrain uncertainty of 0.1\%, motivated by present limits on how well twist angle can be determined from transport measurements, how uniform the ``best'' samples appear to be over micron-scale lengths~\cite{uri_mapping_2020, pierce_unconventional_2021, lu_superconductors_2019, diez-merida_high-yield_2024}, and how small a change in these parameters influences electronic properties seen in transport.

\begin{figure*}
    \centering
    \includegraphics[width=\linewidth]{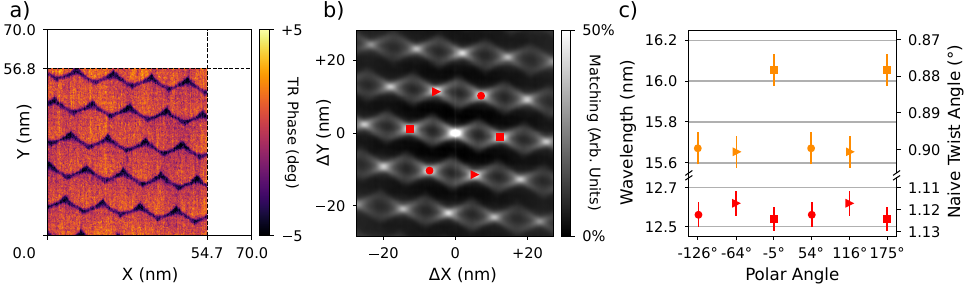}
    \caption{
        \label{Fig1}
        \textbf{Scan area correction and measurement of moir\'e wavelength.}
        (a) TFM phase image of TBG. The frame outline represents the intended scan area (70 nm x 70 nm). The black dashed lines show the smaller and non-square (54.7 nm x 56.8 nm) extent of the actual scan area recorded by X/Y sensors which track the motion of piezos that drive the scanning probe. Further analysis steps are based on these sensor-corrected coordinates. (b) Two-dimensional autocorrelation (AC) map using the sensor-corrected coordinates. The red symbols mark the average moir\'e lattice vectors. (c) The moir\'e wavelength and corresponding naive twist angle of the AC peaks using the intended (orange) or sensor-corrected (red) coordinates. Points are labeled by their polar angle with respect to the X-axis. The errors in the wavelength are set by the pixel size in the maps.
        }
\end{figure*}

The relative twist angle between layers can be extracted from the moir\'e lattice vectors \cite{kerelsky_maximized_2019, kogl_moire_2023, halbertal_extracting_2022}. In previous work we have shown that the lattice vectors can be precisely measured from TFM images given the technique's high contrast and lateral spatial resolution \cite{pendharkar_torsional_2024}. Since then we have found that the lattice vectors extracted from successive scans of the same area are generally different, indicating the accuracy of these measurements does not match their precision. We have revisited our analysis to understand and overcome such scan-to-scan discrepancies.

In analyzing scanning probe data, the pixels in the image are typically assumed to fall on an evenly-spaced square grid (the ``intended grid''). This condition is often roughly enforced by closed-loop feedback on X and Y position during scanning \cite{barrett_optical_1991}. We instead choose to scan in open loop, which allows us to scan at faster line rates (up to 8-24 Hz) and in turn acquire more extensive statistics and survey more regions of our samples. Without feedback, the actual tip position during piezo-driven scanning differs from the corresponding position on the intended grid. This deviation can differ from scan to scan, contributing to our observed discrepancy in extracting lattice vectors \cite{barrett_optical_1991, yothers_real-space_2017}. To accurately assign real-space locations to the pixels in an image, we record the same X/Y position sensor data that would commonly be used for closed-loop control, then use it in post-processing to construct an X/Y coordinate grid.

After correcting with X/Y position sensors, moir\'e lattice vectors extracted from successive scans of alternating slow scan direction show systematic errors. Lattice vectors extracted from scans with the same slow scan direction are internally consistent, but comparison of the vectors between the two slow scan directions show differences up to a few percent. This systematic pattern is consistent with slow, constant-velocity relative drift between sample and probe, which should produce a Doppler-like shift in the extracted periodicity of the moir\'e when scanning along or against the drift direction \cite{yothers_real-space_2017, rahe_vertical_2010, salmons_correction_2010}. To correct for relative drift, we analyze multiple scans of the same region to extract the drift velocity. We then apply a drift-velocity-dependent transformation to the extracted lattice vectors to undo the distortion produced by that drift. We further validate this analysis by applying it to (1) scans with different line-scan rates and (2) scans with slow scan along an axis perpendicular to that for the original batch. If these scans are taken in close succession, all yield a single consensus set of lattice vectors upon correcting for a single vector drift velocity that does not change between scans.

\begin{figure*}
    \centering
    \includegraphics[width=\linewidth]{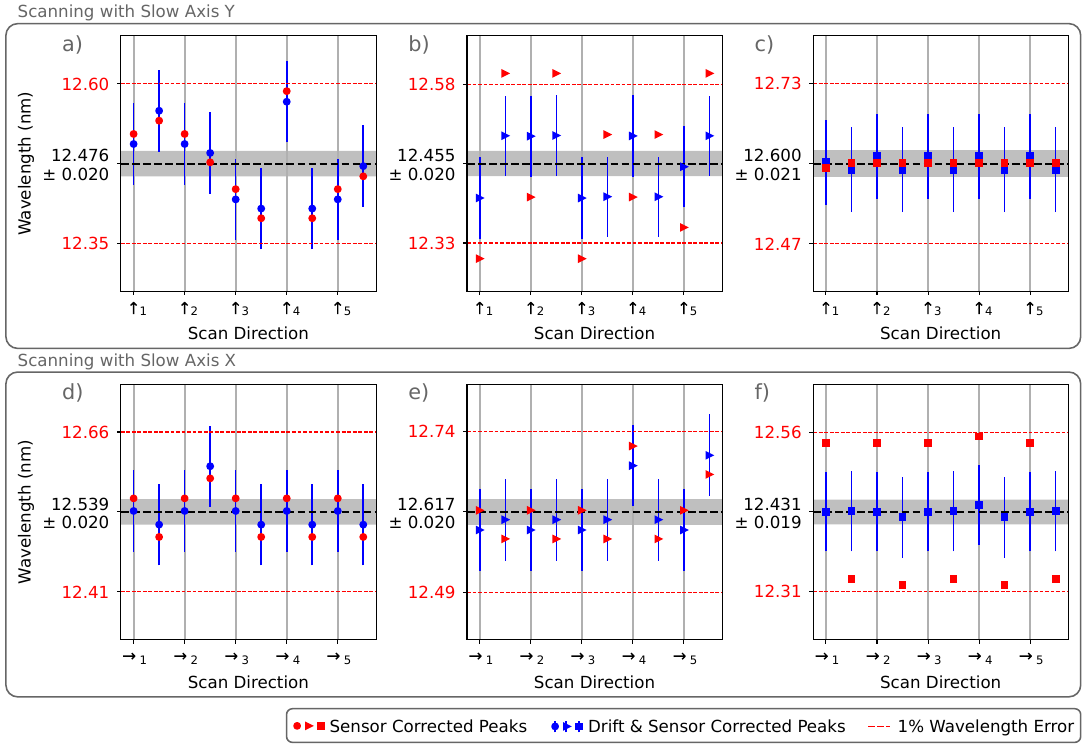}
    \caption{
        \label{Fig2}
        \textbf{Correction of distortion from relative sample-probe drift.} 
         a)-c) Wavelengths for different moir\'e lattice vectors over ten successive scans of alternating scan direction, after X/Y sensor data corrections, with the slow axis being the Y direction. Refer to Figure \ref{Fig1} for the corresponding lattice vector of the marker symbol. Points on vertical gray lines represent scans with slow scan direction of positive slow axis; scans with a scan direction of negative slow axis are interleaved between. Before drift correction (red markers), the wavelengths of lattice vectors marked by triangles in Fig~\ref{Fig1} systematically differ between forward and backward scans (panel b). After drift correction (blue markers), this asymmetry is reduced. Error bars include the error from finite pixel size in our original image. For drift-corrected data, uncertainty in drift velocity is also incorporated through Monte Carlo simulation (see supplementary material). The error bars for the red points are omitted for visibility, but are similar to those for sensor-corrected data in Figure \ref{Fig1}. d)-f) Same as above except the slow axis being the X direction. The data were taken immediately after a)-c). Prior to drift correction, the square points (panel f) rather than triangles now display Forward/Backward asymmetry. For each slow scan direction, the drift-corrected wavelength for each lattice vector from each of ten scans falls within the horizontal dashed red lines, which denote a +/-1\% error around the central value of this wavelength. See supplementary material for significance of +/-1\% error.
        }
\end{figure*}

A TFM image of TBG that has been corrected with X/Y position sensors is shown in Figure \ref{Fig1}a. The scan was intended to be a 70 nm square frame, but the measured scan dimensions are about 20$\%$ smaller than intended and rectangular rather than square. The effect of this correction on the extracted twist angle can be seen in Figure \ref{Fig1}b-c which shows our extraction and analysis of moir\'e lattice vectors. To extract the lattice vectors, we employ a real-space analysis using autocorrelations (AC) (see supplementary material). Figure \ref{Fig1}b shows a two-dimensional AC of the image using the sensor-corrected coordinates. The red markers denote peaks in the AC which represent the average moir\'e lattice vectors. From these vectors we calculate the moir\'e wavelength -- the length of the vector -- and convert the wavelength to a ``naive'' twist angle, for now neglecting strain and treating each vector as independent from the others. The moir\'e wavelengths and twist angles are shown in Figure \ref{Fig1}c for the uncorrected (orange markers) and sensor-corrected (red markers) images. The scan area correction results in an approximate 0.2° change in extracted twist angle. 

The uncertainty due to relative drift between sample and probe can be seen by keeping track of the slow scan direction. We refer to scans which have a slow scan direction of positive slow axis as Forward (F) and slow scan direction of negative slow axis as Backward (B). Shown in Figure \ref{Fig2} are two back-to-back sets of 10 successive (F, B, F, ...) scans. Each scan is acquired at a line-scan rate of 8 Hz with 512x512 pixels. The moir\'e wavelengths of the three unique moir\'e lattice vectors are plotted as a function of the scan direction. With only sensor correction (red markers), we see an asymmetry between Forward and Backward scans in Figure \ref{Fig2}b, f. In general, whether a lattice vector shows asymmetry or not depends on the polar angle of the lattice vector, direction of drift, and scan slow axis. This is why the triangle markers show asymmetry in Figure \ref{Fig2}b but not in Figure \ref{Fig2}e, and vice versa for the square markers.

\begin{table*}
\centering
\begin{tabular}{c|c| c|c|c}
     & Slow axis  & Line scan rate (Hz) &Drift along X (nm/min) &Drift along Y (nm/min) \\ \hline
     Set 1 & Y & 8 & 0.532 +/- 0.055 & -0.270 +/- 0.050 \\ \hline
     Set 2 & X & 8 & 0.480 +/- 0.053 & -0.140 +/- 0.054 \\ \hline
     Set 3 & X & 4 & 0.498 +/- 0.034 & -0.034 +/- 0.036 \\ \hline
\end{tabular}

    \caption{Estimated drift velocities across three successive sets of measurements with different slow axes. Sets 1 and 2 (shown in Figure \ref{Fig2}) contain 10 successive scans and are acquired with a line scan rate of 8 Hz. Set 3 contains six scans and is acquired with a line scan rate of 4 Hz. Each set of measurements takes roughly 10 minutes. Each scan contains 512x512 pixels.}
    \label{Drift_Table}
\end{table*}

Constant-velocity relative drift linearly transforms the true lattice vectors into the distorted lattice vectors we extract. Correcting for drift distortion requires us to apply the inverse transformation on our extracted vectors. See supplementary material for derivation of this transformation, as well as details on how we extract the relative drift velocity. Table \ref{Drift_Table} shows our estimated drift velocities for the two sets of measurements shown in Figure \ref{Fig2}, as well as a third set taken immediately after the second with a different line scan rate. The estimated drift velocities are on the order of hundreds of pm/min and are consistent across the three sets which in total took roughly 30 minutes to acquire. The consistency of the velocities suggests that we may treat established drift values as constant when correcting future scans close in time. However, drift calibrations should be performed periodically to verify trends in the drift and to get the most accurate estimation.

The results of drift correction on our extracted lattice vectors are shown in blue in Figure \ref{Fig2}. The Forward/Backward asymmetry is slightly reduced in Figure \ref{Fig2}b and nearly eliminated in Figure \ref{Fig2}f. After drift correction, the average wavelengths of corresponding peaks nearly match between data sets with orthogonal slow axes but have a small deviation by 60 to 200 pm, comparable to the 100 pm pixel size. After drift correction, we extract the TBG twist angle, uniaxial heterostrain magnitude and strain angle by treating the lattice vectors collectively rather than individually and using vector displacements rather than wavelength. These parameters are extracted by applying rotation and uniaxial strain transformations on ideal graphene layers to calculate the expected moir\'e lattice vectors \cite{kerelsky_maximized_2019}. An optimization of these transformations is performed until the expected vectors agree with our measured vectors (see supplementary material). 

\begin{figure*}
    \centering
    \includegraphics[width=\linewidth]{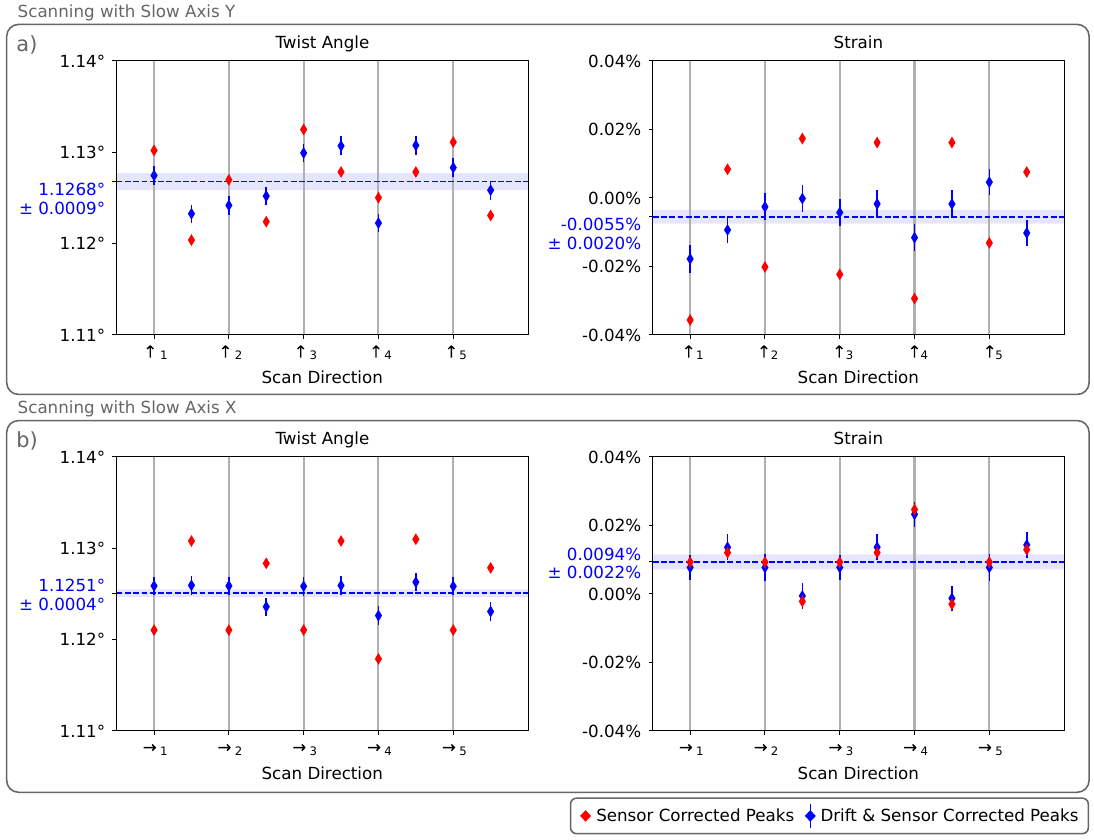}
    \caption{
        \label{Fig3}
        \textbf{Extraction of TBG twist angle and uniaxial heterostrain.} 
        Results of optimization for (a) slow axis Y and (b) slow axis X, with sensor corrections (red) and drift and sensor corrections (blue). The dashed blue line is the mean value of the parameter after drift correction. The reported error of the mean is statistical uncertainty. The error bars on points are obtained by repeating the optimization using Monte Carlo simulation from a distribution around our extracted moir\'e lattice vectors, then fitting a Gaussian to the resulting distribution of parameters. The values of the parameters extracted from scans with the two different slow axes are consistent after drift correction. 
        }
\end{figure*}

The optimized TBG twist angle, uniaxial heterostrain and strain angle are shown in Figure \ref{Fig3}. After drift correction, we found that the average twist angle of our images is $1.126$°, close to the TBG magic-angle. Both the statistical uncertainty within each data set and the difference between the two sets with perpendicular slow axis directions are on the order of a thousandth of a degree. This region of our sample has an average uniaxial heterostrain magnitude of less than a hundredth of a percent; for such weak heterostrain, we cannot confidently extract the angle along which the strain is applied. The lack of heterostrain may be a result of our scanning in a region which lacks bubbles and wrinkles. See supplementary material for a discussion of strain angle and a larger-area scan showing the landscape of bubbles and wrinkles in our sample.

In post-processing, we are able to correct scanning probe images to accurately extract moir\'e lattice vectors with less than 1\% error. This enables us to extract twist angles near TBG magic-angle with uncertainty on the order of a thousandth of a degree. Were we to analyze images of TBG with a different twist angle, our uncertainty could be interpreted as a maximum (minimum) uncertainty for twist angles smaller (larger) than TBG magic-angle. Figure S5 (supplementary material) shows the naive twist angle of TBG as a function of the moir\'e wavelength. The error bounds on twist angle correspond to a scenario where there is a flat percentage error in measuring wavelength. In general, the uncertainty in twist angle increases as the extracted wavelength becomes smaller. 

In this work we analyze images tens of nanometers in size, spanning a few moir\'e periods across each image. For comparison, electrical transport measurements probe the properties between contact pairs which are typically spaced a micron apart. Depending on the twist angle between VdW layers, a few to hundreds of moir\'e periods can be contained between the contacts. Quantitative analysis of scanning probe images spanning larger regions \cite{mcgilly_visualization_2020, kapfer_programming_2023} -- one to a few microns across -- could provide useful structural information to complement electrical transport measurements. For example, uniaxial heterostrain was recently shown to be responsible for dramatic and novel electrical properties of a TBG sample~\cite{finney_unusual_2022}. In that study, heterostrain was not intentionally introduced, and no direct measure of heterostrain was available. Instead the presence and magnitude of uniaxial heterostrain was inferred based on an excellent match of theoretical calculations to transport data \cite{wang_unusual_2023}. The analysis protocol described here is usable on larger length scales, provided the moir\'e can be resolved and the relative drift velocity can be estimated. Applying this protocol to larger-area scanning probe images of open-face stacks prior to encapsulation would provide a direct measure of heterostrain in a moir\'e and could even be used to select a particular region of a moir\'e on which to form an electrical device. 

Application of the analysis can also be extended to multi-layer moir\'e heterostructures, provided multiple moir\'es can be imaged in a single sample \cite{pendharkar_torsional_2024, huang_imaging_2021, turkel_orderly_2022}. For example, images of TBG with aligned hBN can be used to identify the relative twist angles between graphene and hBN layers, to screen for samples which may host the quantum anomalous Hall effect \cite{sharpe_emergent_2019, serlin_intrinsic_2020, shi_moire_2021}. Finally, this analysis protocol could also provide rapid feedback for developing stacking processes for improved structural control and uniformity of moir\'es, especially in the context of robotic stacking in vacuum \cite{mannix_robotic_2022, wang_clean_2023}.

In conclusion, we have described a protocol to accurately extract moir\'e lattice vectors from scanning probe images. This protocol is tested by analyzing successive scans of nearly the same area of near-magic-angle twisted bilayer graphene. Systematic errors are first corrected by using position sensors to define an accurate coordinate grid, and then by determining and accounting for a slow relative drift between sample and probe. Finally, we extract twist angle and uniaxial heterostrain using the the full set of moir\'e lattice vectors. The statistical uncertainty in twist angle extracted from an ensemble of ten scans is less than $0.001$°, and the uncertainty in strain is $0.002$\%. Comparing between scans performed with slow axis in two perpendicular directions, these values are consistent, except that the extracted strain differs by $\sim0.01$\%, reflecting a small systematic error we have not yet identified.

\section{Sample preparation \& AFM measurements}
Our sample was prepared in air using polymer stamps from \cite{mannix_robotic_2022}. The sample was imaged using the TFM protocol in \cite{pendharkar_torsional_2024} where extensive details regarding the technique and the measurements performed have been provided.

All AFM measurements shown as part of this work were performed at Stanford university in a shared facility Bruker Dimension Icon AFM equipped with NanoScope V electronics and software version 9.40 (Windows 7) and (after an update) version 9.70 (Windows 10, 64-bit). A Standard Operating Procedure (SOP) for Torsional Force Microscopy is available at \cite{pendharkar_torsional_2024}, to aid in the reproduction of these results and a procedure for analysis of AFM results is provided in the supplementary material.

\section{Supplementary material}
See the supplementary material for details on pre-processing of X/Y position sensors + TFM images, a large-area survey scan of the sample, derivation of the drift transformation matrix, estimation of relative drift velocities, extraction of relative twist angles and heterostrain, propagation of uncertainties and additional data for Set 3 of Table \ref{Drift_Table}.

\section{Acknowledgments}
We thank Xiaoyu Wang, Christina Newcomb, Bede Pittenger, Peter De Wolf, Javier Sanchez-Yamagishi, Andrew Barabas, Lloyd Bumm, Maelle Kapfer, and Cory Dean for fruitful discussions.

\section{Funding}
Sample preparation, measurements, and analysis were supported by the US Department of Energy, Office of Science, Basic Energy Sciences, Materials Sciences and Engineering Division, under Contract DE-AC02-76SF00515. Development of tools for robotic stacking of 2D materials were supported by SLAC National Accelerator Laboratory under the Q-BALMS Laboratory Directed Research and Development funds. All AFM imaging reported here was performed at the Stanford Nano Shared Facilities (SNSF), and stamps for stacking were prepared in Stanford Nanofabrication Facility (SNF), both of which are supported by the National Science Foundation under award ECCS-2026822.  M.P. acknowledges partial support from a Stanford Q-FARM Bloch Postdoctoral Fellowship. D.G.-G. acknowledges support for supplies from the Ross M. Brown Family Foundation and from the Gordon and Betty Moore Foundation’s EPiQS Initiative through grant GBMF9460. The EPiQS initiative also supported a symposium of early career researchers which enabled feedback from the community on this work during its development. M.H. acknowledges partial support from the Department of Defense through the Graduate Fellowship in STEM Diversity program. K.W. and T.T. acknowledge support from the JSPS KAKENHI (Grant Numbers 21H05233 and 23H02052) and World Premier International Research Center Initiative (WPI), MEXT, Japan.

\section{Data Availability}
The data that support the findings of this study are openly available in Stanford Digital Repository at https://doi.org/10.25740/ym579qg7863.

\section{Duration and Volume of Study}
TFM data used as part of this work were acquired between April 2023 (when we started recording position sensor data) and March 2024. Development of the analysis protocol described here was concluded in May 2024.

\section{Author Declarations: Conflict of Interest}
M.A.K. served as a member of the Department of Energy Basic Energy Sciences Advisory Committee until December 2023. Basic Energy Sciences provided funding for this work.
M.A.K. served as an independent director on the board of Bruker Corporation until May 2023. All data shown were taken on a Bruker Dimension Icon AFM at Stanford University.

\section{Author Contributions}
\textbf{Steven Tran:} Data curation (equal); Formal analysis (lead); Investigation (equal); Software (equal); Visualization (equal);  Writing – original draft (equal); Writing – review $\&$ editing (equal). \textbf{Jan-Lucas Uslu:} Formal analysis (supporting); Software (equal); Visualization (equal). \textbf{Mihir Pendharkar:} Data curation (equal); Formal analysis (supporting); Investigation (equal); Writing – original draft (supporting). \textbf{Joe Finney:}  Resources (equal); Writing – original draft (supporting). \textbf{Aaron L. Sharpe:} Resources (equal); Writing – original draft (supporting). \textbf{Marisa L. Hocking} Resources (equal). \textbf{Nathan J. Bittner:} Resources (equal). \textbf{Takashi Taniguchi:} Resources (equal). \textbf{Kenji Watanabe:} Resources (equal). \textbf{Marc A. Kastner} Funding acquisition (equal); Supervision (equal); Writing – original draft (supporting); Writing – review $\&$ editing (equal). \textbf{Andrew J. Mannix:}  Funding acquisition (equal); Supervision (equal); Writing – original draft (supporting); Writing – review $\&$ editing (equal). \textbf{David Goldhaber-Gordon:} Conceptualization (lead); Funding acquisition (equal); Supervision (equal); Writing – original draft (supporting); Writing – review $\&$ editing (equal).

\bibliographystyle{apsrev4-2}
\bibliography{references.bib}

\beginsupplement
\justifying

\section{Pre-processing of scanning probe data}

\subsection{(Torsional resonance channels)}
Information about the real space structure of the moir\'e superlattice are captured in the torsional resonance (TR) amplitude and phase channels of torsional force microscopy (TFM). If the raw data has high signal-to-noise with little background, we prefer to work directly with the raw data that has an average value plane subtracted from it. Figure 1a of the main text is an example of the raw data with an average plane subtracted from it. If background subtraction is necessary for autocorrelation or fast Fourier transform analyses, we perform a fit of a 2D polynomial background (up to third degree polynomial in X and Y) to remove it. We avoid working with a line-by-line background subtraction (fitting each line to a polynomial) because information about the moir\'e will be lost if any of its features align with a scan line.

\begin{figure}[h!]
    \centering
    \includegraphics[width=\linewidth]{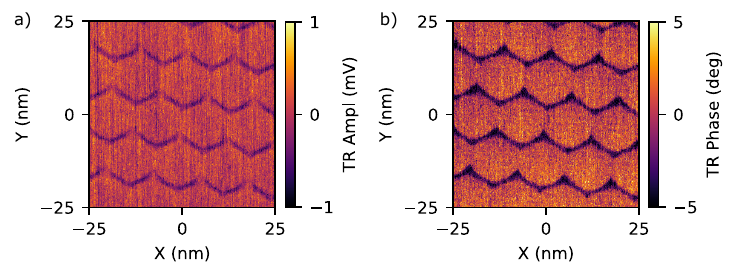}
    \caption{
        \label{Supp_Fig1}
        \textbf{Torsional resonance amplitude and phase.} 
        Example TR amplitude (a) and phase (b) channels from TFM. The raw data with a plane of the average value subtracted are shown. 
        }
\end{figure}

\clearpage
\subsection{(X/Y position sensors)}
\begin{figure}[h!]
    \centering
    \includegraphics[width=\linewidth]{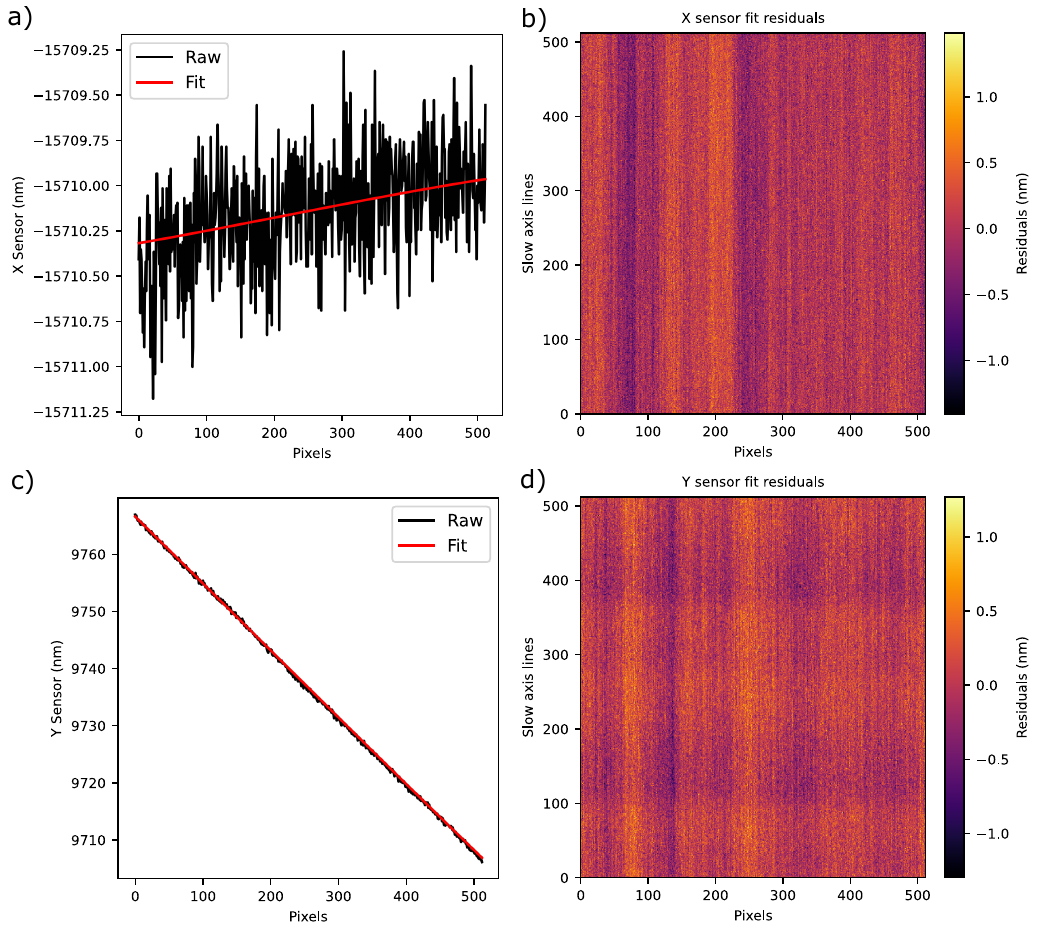}
    \caption{
        \label{Supp_Fig2}
        \textbf{Raw X and Y position sensors with the residuals of 2D polynomial fit.} 
        (a) X position sensor and (c) Y position sensor for a line of the scan (with fast axis Y and slow axis X). The black curve is the raw sensor data which shows hundreds of picometers of noise. The red curve is the linear fit to the sensor data. (b) and (d) show the residuals of the 2D polynomial fit to the sensor data.
        }
\end{figure}

The raw data from position sensors have noise on the order of hundreds of picometers which lead to unrealistic real space placements of pixels when constructing an X/Y coordinate grid using the raw data directly (see Figure \ref{Supp_Fig2}a). Furthermore, the pixels on the raw grid are not necessarily equally spaced, causing issues for autocorrelation or fast Fourier transform analyses which expect equally spaced pixels. To remove the noise and obtain equally spaced pixels, we make the assumption that the motion of the scanning probe is accurately described by a linear fit of X and Y sensor data. We independently fit the X and Y sensors to a 2D first degree polynomial and construct an X/Y coordinate grid from the fitted sensor data.

\begin{figure}
    \centering
    \includegraphics[width=0.9\linewidth]{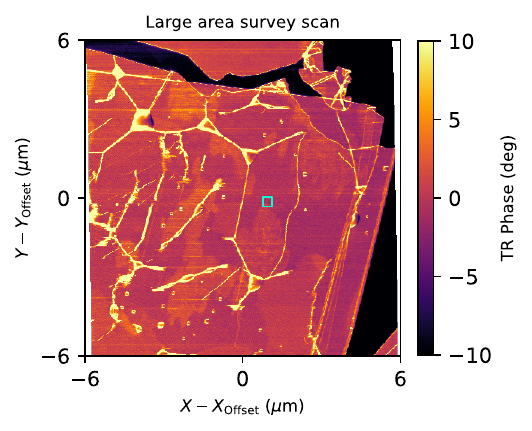}
    \caption{
        \label{Supp_Fig3}
        \textbf{Large area scan showing sample bubbles and wrinkles.} 
        Sensor-corrected large-area survey scan. The intended scan size was 12 um x 12 um. The blue square corresponds to the approximate region the data in the main text were taken at. Our scan region is not fully enclosed by wrinkles, nor is it close to bubbles. This may explain the lack of heterostrain in our analysis. The straight vertical line to the left of the blue square is a real wrinkle in the sample, the curvy line to the right corresponds to a dust particle that is being dragged around by the probe throughout the scan.
        }
\end{figure}

\clearpage
\section{Analysis of images: autocorrelations and fast Fourier transforms}
We use two-dimensional autocorrelations (AC) as our primary tool for analyzing our TFM images and extracting the average moir\'e lattice vectors. The AC displaces an image with respect to itself, takes the product of overlapping pixels and then sums over all the products. When ``identical'' features overlap, there is a peak in the AC spectrum. For images of moir\'es, the peaks closest to zero displacement correspond to when neighboring moir\'e unit cells overlap, which are the moir\'e lattice vectors by definition. When the AC is performed with images that contain more than three unit cells, the peaks represent the average moir\'e lattice vectors because we are simultaneously determining the lattice vectors from every unit cell present in the image. To get the ``local'' lattice vectors at a given point in the image, the data should be cropped around the point to include only neighboring unit cells. A fast Fourier transform (FFT) analysis can also be employed. We prefer AC for our analysis because our images have high contrast which allows us to work in real space directly. Furthermore, the FFT method can run into issues when there only a few periods in the image, whereas the AC does not. In Figure \ref{Supp_Fig3} we compare the extracted moir\'e wavelengths from the AC versus FFT, which agree quite well (within a few hundred picometers.) To get the FFT, a Hanning window and zero-padding were applied to the input image to get reduce FFT artifacts and to increase the resolution of the FFT.

\begin{figure}[h!]
    \centering
    \includegraphics[width=0.95\linewidth]{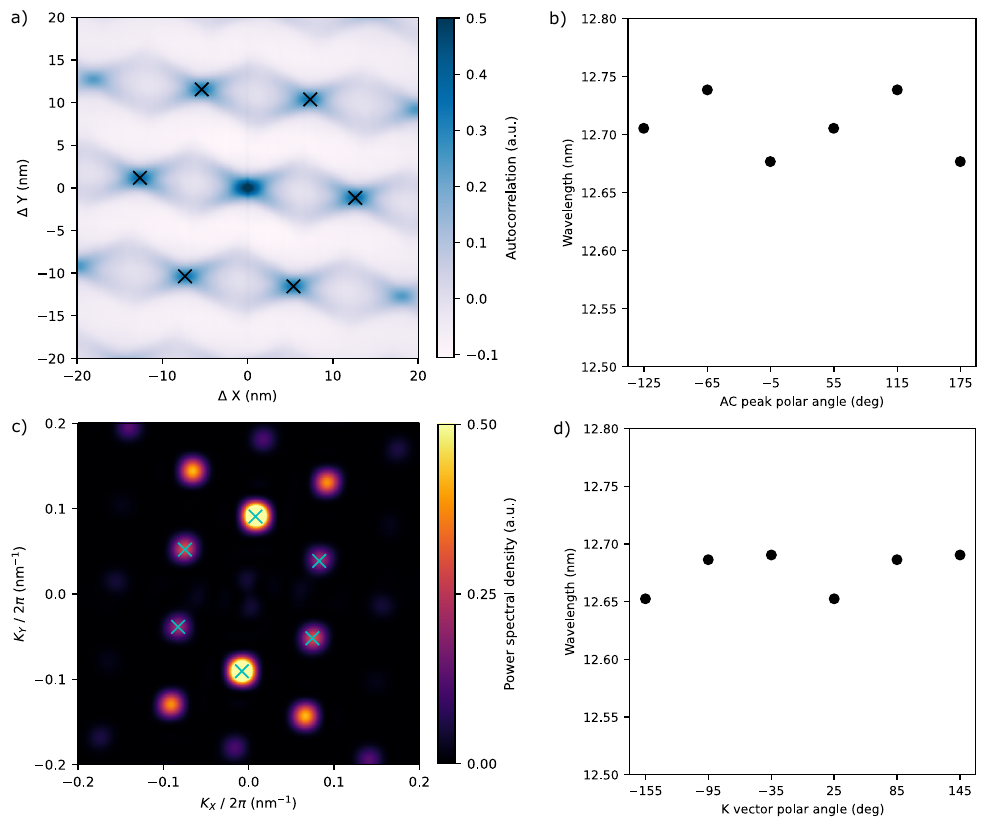}
    \caption{
        \label{Supp_Fig4}
        \textbf{Comparison of autocorrelation and fast Fourier transform analysis.} 
        (a) A two-dimensional autocorrelation map for the image shown in Figure \ref{Supp_Fig1}. Black crosses denote the average moir\'e lattice vectors. (b) The wavelength (length of the vector) for the various lattice vectors, labeled by their polar angle with respect to the X-axis. (c) A two-dimensional FFT for the same input image as (a). Shown is the power spectral density (square of the FFT magnitude).  The blue crosses correspond to the reciprocal moir\'e lattice vectors $\vec{K}_i$. (d) The corresponding real space wavelength of the reciprocal vector (approximated by 2/$\sqrt{3}$/$|\vec{K}_i|$). Both the AC and FFT were normalized by the maximum value of the spectrum. 
        }
\end{figure}

\clearpage 

\section{Propagation of moir\'e wavelength error to naive twist angle}
\begin{figure}
    \centering
    \includegraphics[width=\linewidth]{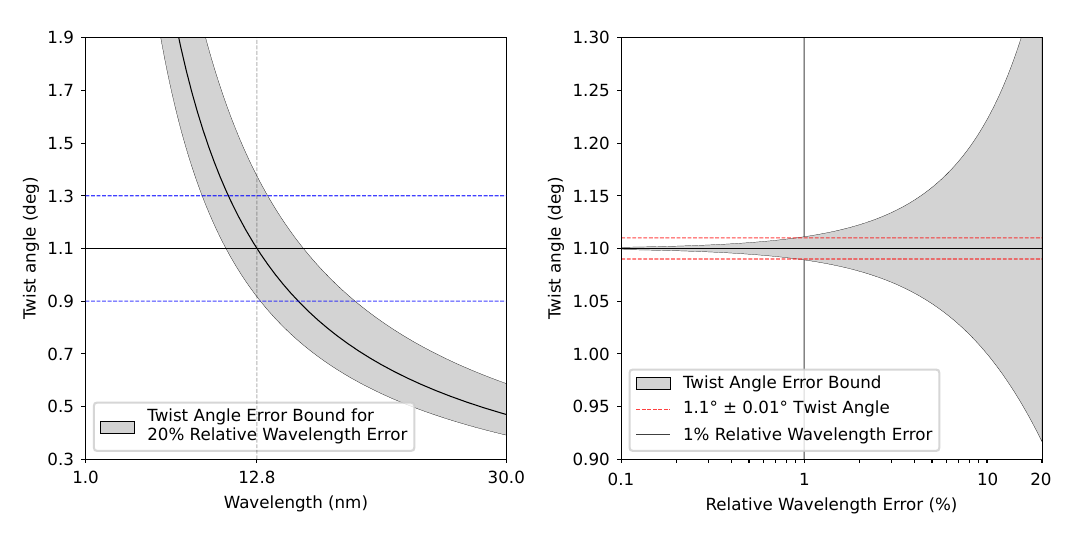}
    \caption{
        \label{Supp_Fig5}
        \textbf{Propagation of moir\'e wavelength uncertainty to twist angle.} 
        (a) Twist angle of TBG as a function of the moir\'e wavelength. The shaded regions represent the bound on twist angle error resulting from a 20$\%$ relative error in measuring wavelength. Dashed blue lines are a guide to the eye showing an approximate uncertainty of $\pm$0.2$^\circ$ in determining TBG magic-angle (1.1$^\circ$) with a 20$\%$ relative error. (b) Scaling of the twist angle error near magic angle as a function of relative error in wavelength. The dashed horizontal red lines show an uncertainty of $\pm$0.01$^\circ$. The vertical black line denotes the maximum wavelength error (1$\%$) consistent with this. 
        }
\end{figure}

Shown in Figure \ref{Supp_Fig5} (left panel) is the naive twist angle for TBG as a function of moir\'e wavelength when there is fixed 20$\%$ error in measuring the wavelength (see the main text for definition of naive twist angle). Generally, the uncertainty in twist angle increases as the measured moir\'e wavelength decreases. If we focus on twist angles around the TBG magic angle (1.1$^\circ$), a 20$\%$ fixed error leads to an uncertainty on the order of +/- 0.2°. If the goal is to determine the magic angle to within a hundredth of a degree, an error in the measurement of the moire wavelength less than 1$\%$ is required (right panel of Figure \ref{Supp_Fig5}(b)).

\clearpage
\section{Derivation of the linear transformation due relative drift between probe and sample}
Before we proceed with the derivation, we will need to make assumptions about the dynamics of the scanning probe to model our experiment. We will assume a fast scan axis of Y, a slow scan axis of X, and a scan area with dimensions ($L_x \times L_y$) that is centered around the sample origin $O = (0,0)$. The derivation swapping the scan axes will be identical. Next, we need to estimate the fast and slow probe velocities. In an actual experiment, these velocities are dependent on the scan dimensions and two software inputs: the line scan rate ($f$) and number of lines ($N$) in an image. We choose the fast probe velocity to be $\vec{V}_{fast} = (0, V_{fast})$ ($V_{fast} = 2 L_y f$) and the slow probe velocity to be $\vec{V}_{slow} = (V_{slow}, 0)$ ($V_{slow} = L_x f/ N$). These velocity vectors point from the starting point of a scan, which we discuss in the next paragraph. Because our fast and slow probe velocities are chosen to align perfectly along the fast and slow axes, the scan area will be rectangular by construction. This choice makes our calculations simpler, but the scan area in an actual experiment may be a parallelogram because probe velocities are typically not perfectly aligned with the scan axes. We find the rectangular scan area to be a good approximation even if the scan area is slightly parallelogram. For further simplication, assume a square scan area ($L_x = L_y = L$). The last ingredient we need is a distinction between Forward (F) and Backward (B) scan directions. In an actual measurement, the difference between Forward and Backward scans is the starting point of the scan, and the sign of the slow probe velocity. For a Forward scan, we have slow probe velocity +$\vec{V}_S$ with a starting point of $O_F = (-L/2, -L/2)$. For a Backward scan, we have slow probe velocity -$\vec{V}_S$ with starting point of $O_B = (L/2, -L/2)$. For the derivation swapping the fast and slow axes, we would keep $O_F$ the same but $O_B = (-L/2, L/2)$. To keep track of the sign of the slow velocity with respect to scan direction, we redefine $\vec{V}_S = (C_A \cdot V_S, 0)$ ($V_S = L_x f/ N$) where $C_A = 1$ if $A$ = F, and $C_A = -1$ if $A$ = B. 

Moving on to the derivation, let us assume there is a feature of interest on our sample at coordinates $R = (R_x, R_y)$. We define our true lattice vector $\vec{R}$ as the vector which points from $O$ to the feature ($\vec{R} = R - O = (R_x, R_y)$). The next vector we define is $\vec{T}_A(t)$, which is a vector which points from the scan starting point $O_A$ to the position of the probe as a function of time in the scan. The position of the probe is given by the fast and slow probe velocities, and two times: the ``slow" time $t_{slow} = n/f$ where $n$ is the number of lines that have been scanned, and the ``fast" time $t_{fast}$ which is the time spent on the current line. $t_{slow}$ will determine the vector displacement along the slow axis, and $t_{fast}$ will determine the vector displacement along the fast axis. We can then write $\vec{T}_A(t = t_{slow} + t_{fast}) = O_A + \vec{V}_{slow} \cdot t_{slow} + \vec{V}_{fast} \cdot t_{fast}$. The vector $\vec{T}_A$ can be used to calculate the time the scanning probe would arrive at $O$ or $R$, which we call $t_O$ and $t_R$, respectively. This calculation can be done by solving,

\begin{equation}
    \label{O_time}
    \begin{aligned}
        \vec{T}_A(t_O = t_{O, slow} + t_{O, fast})  = O \\
         \vec{V}_{slow} \cdot t_{O,slow} + \vec{V}_{fast} \cdot t_{O,fast} + O_A = O \\
         (C_A V_{slow} \cdot t_{O,slow}, 0) + (0, V_{fast} \cdot t_{O, fast}) = O - O_A. \\
    \end{aligned}
\end{equation}

\begin{equation}
    \label{R_time}
    \begin{aligned}
        \vec{T}_A(t_R = t_{R, slow} + t_{R, fast})  = R \\
         \vec{V}_{slow} \cdot t_{R,slow} + \vec{V}_{fast} \cdot t_{R,fast} + O_A = R \\
         (C_A V_{slow} \cdot t_{R,slow}, 0) + (0, V_{fast} \cdot t_{R, fast}) = R - O_A.\\
    \end{aligned}
\end{equation}

To derive the linear transformation due to relative drift, we need to calculate how $\vec{R}$ is perceived by the scanning probe in the presence of relative drift between probe and sample. To introduce drift dynamics we assume all the relative drift is on the sample side with drift velocity $\vec{D} = (D_x, D_y)$. This means the coordinates of $O$ and $R$ are changing over time according to $O^*(t) = O + \vec{D}t$ and $R^*(t) = R + \vec{D}t$, where $t$ is the elapsed time since the start of the scan. The transformed $\vec{R}$ is given by $\vec{R}^* = R^*(t_R) - O^*(t_O)$. With the introduction of relative drift, Equations \eqref{O_time} and \eqref{R_time} become

\begin{equation}
    \label{Drift_O_time}
    \begin{aligned}
         (C_A V_{slow} \cdot t_{O,slow}, 0) + (0, V_{fast} \cdot t_{O, fast}) = O + \vec{D} \cdot (t_{O, slow} + t_{O, fast}) - O_A,
    \end{aligned}
\end{equation}

\begin{equation}
    \label{Drift_R_time}
    \begin{aligned}
         (C_A V_{slow} \cdot t_{R,slow}, 0) + (0, V_{fast} \cdot t_{R, fast})  = R + \vec{D} \cdot (t_{R, slow} + t_{R, fast}) - O_A, 
    \end{aligned}
\end{equation}
We will make the following approximation to make the algebra simpler by decoupling $t_{slow}$ and $t_{fast}$ in one equation, $\vec{D} \cdot (t_{slow} + t_{fast}) = (D_x \cdot (t_{slow} + t_{fast}), D_y \cdot (t_{slow} + t_{fast})) \approx (D_x \cdot t_{slow}, D_y \cdot (t_{slow} + t_{fast})$). Physically, $t_{slow}$ corresponds to the time when the probe catches up to the feature along the slow axis and is along the same line. At this time, the feature would have drifted along the slow axis by $D_x \cdot t_{slow}$ (recall X is our slow axis). Once the probe arrives at the same line, it catches up to the feature along the fast axis in some time $t_{fast}$. The original expression before the approximation $D_x \cdot (t_{slow} + t_{fast})$ takes into the account the extra drift along the slow axis in the time $t_{fast}$. Since $t_{fast}$ is counted from the start of the current line, we assume it to be negligible when compared to $t_{slow}$, so we ignore the extra contribution. We will proceed with the derivation assuming the Forward scan direction ($A$ = F). Solving Equation \ref{Drift_O_time} first,

\begin{equation*}
    \begin{aligned}
        (C_F V_{slow} \cdot t_{O,slow}, 0) + (0, V_{fast} \cdot t_{O, fast}) = O + \vec{D} \cdot (t_{O, slow} + t_{O, fast}) - O_F, \\
        (V_{slow} \cdot t_{O,slow}, 0) + (0, V_{fast} \cdot t_{O, fast}) = (D_x \cdot t_{O, slow} + L/2, D_y \cdot (t_{O, slow} + t_{O, fast}) + L/2).
    \end{aligned}
\end{equation*}
By comparing components we get two equations

\begin{equation*}
    \begin{aligned}
        V_{slow} \cdot t_{O,slow} = D_x \cdot t_{O, slow} + L/2 \Rightarrow t_{O, slow} =  \frac{1}{V_{slow} - D_x} \frac{L}{2}.\\
        V_{fast} \cdot t_{O, fast} =  D_y \cdot (t_{O, slow} + t_{O, fast}) + L/2  \Rightarrow t_{O, fast} = \frac{1}{V_{fast} - D_y}  \left[ \frac{D_y}{V_{slow}-D_x} + 1\right] \frac{L}{2}.
    \end{aligned}
\end{equation*}
Doing the same algebra for Equation \ref{Drift_R_time}, 
\begin{equation*}
    \begin{aligned}
        t_{R, slow} = \left(R_x + \frac{L}{2} \right) \frac{1}{V_{slow} - D_x}. \\
        t_{R, fast} = \frac{1}{V_{fast} - D_y} \left[R_y + \frac{D_y}{V_{slow}-D_x} R_x \right] + \frac{1}{V_{fast} - D_y}  \left[ \frac{D_y}{V_{slow}-D_x} + 1\right] \frac{L}{2}.
    \end{aligned} 
\end{equation*}
We can now solve for the transformed $\vec{R}^* = R^*(t_{R, slow} + t_{R, fast}) - O^*(t_{O, slow} + t_{O, fast})$,
\begin{equation*}
    \begin{aligned}
        \vec{R}^* = R + \vec{D} \cdot (t_{R, slow} + t_{R, fast}) - O - \vec{D} \cdot (t_{O, slow} + t_{O, fast}) \\
        \approx (R - O) + (D_x \cdot (t_{R, slow} - t_{O, slow}), D_y \cdot (t_{R, slow} - t_{O, slow} + t_{R, fast} - t_{O, slow})) \\
        = (R_x, R_y) + \left(D_x \cdot \frac{R_x}{V_{slow}-D_x}, D_y \cdot \left[ \frac{R_x}{V_{slow}-D_x} + \frac{1}{V_{fast} - D_y} \left[R_y + \frac{D_y}{V_{slow}-D_x} R_x \right]  \right] \right).
    \end{aligned}
\end{equation*}
This can be re-written as a matrix equation,
\begin{equation}
    \begin{aligned}
        (\vec{R}^*)^T = 
        \begin{bmatrix}
            [1 + D_x/(V_{slow}-D_x)] & 0 \\
            D_y/(V_{slow}-D_x)*[1 + D_y/(V_{fast}-D_y)]& [1 + D_y/(V_{fast} - D_y)]
        \end{bmatrix}
        \vec{R}^T.
    \end{aligned}
\end{equation}
\subsection{Summary}
The linear transformation $T$ due to relative drift (for fast axis Y, slow axis X) and Forward direction,
\begin{equation}
    T (\text{fast - Y, slow - X, Forward}) =         \begin{bmatrix}
            [1 + D_x/(V_{slow}-D_x)] & 0 \\
            D_y/(V_{slow}-D_x)*[1 + D_y/(V_{fast}-D_y)]& [1 + D_y/(V_{fast} - D_y)]
        \end{bmatrix}.
\end{equation}
For fast axis Y and slow axis X, $V_{fast} = 2 L_y f > 0$, $V_{slow} = L_x f/ N > 0$ where $(L_x, L_y)$ are the X and Y dimensions of the scan area, $f$ is the line scan rate and $N$ is the number of lines. $\vec{D} = (D_x, D_y)$ is the relative drift velocity.
Repeating the calculation for Backward direction (essentially $V_{slow} \rightarrow - V_{slow}$), 
\begin{equation}
    T (\text{fast - Y, slow - X, Backward}) =         \begin{bmatrix}
            [1 - D_x/(V_{slow}+D_x)] & 0 \\
            - D_y/(V_{slow}+D_x)*[1 + D_y/(V_{fast}-D_y)]& [1 + D_y/(V_{fast} - D_y)]
        \end{bmatrix}.
\end{equation}
Finally, one can repeat the calculation with fast axis X and slow axis Y to find
\begin{equation}
    T (\text{fast - X, slow - Y, Forward}) =         \begin{bmatrix}
             [1 + D_x/(V_{fast} - D_x)]  & D_x/(V_{slow}-D_y)*[1 + D_x/(V_{fast}-D_x)] \\
            0 & [1 + D_y/(V_{slow}-D_y)]
        \end{bmatrix}.
\end{equation}
\clearpage

\begin{figure}[h!]
    \centering
    \includegraphics[width=\linewidth]{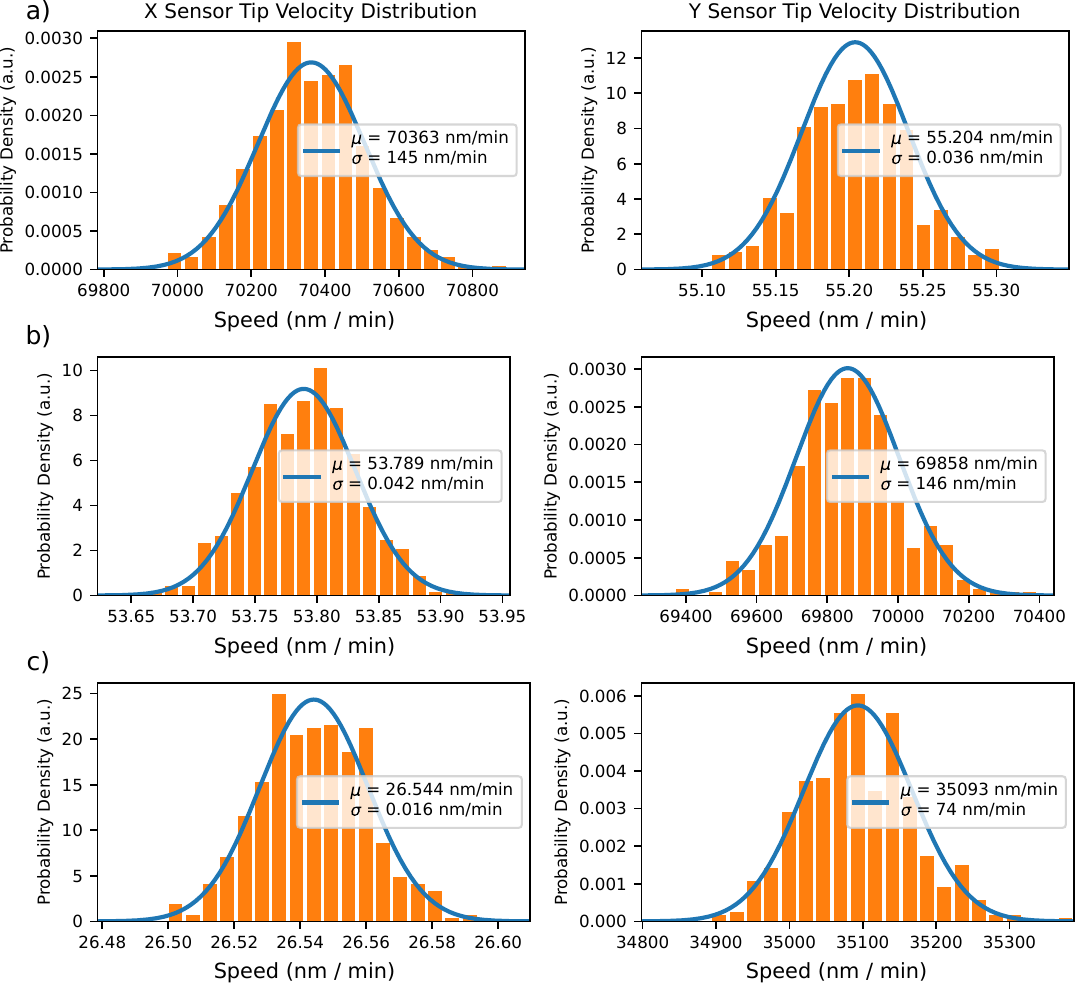}
    \caption{The distribution of tip velocities for the X and Y axis when extracted from the sensor data.
    a shows the speed distribution for the first slow Y Axis scan, b) for the first slow X Axis scan and c) for the first slow X Axis scan of the second set.
}
    \label{Supp_Fig_Vel}
\end{figure}

\section{Explanation of the Tip Velocity estimation}

The fast and slow tip velocities are estimated by fitting a first-order polynomial (mx + b) to each linecut of the X and Y sensors.
For a 512 x 512 scan, this yields 512 values for the slope, which represent the estimated velocities.
The mean ($\mu$) of these values is treated as the nominal tip velocity, and the empirical standard deviation ($\sigma$) is used as the uncertainty.

\begin{align}
\mu &= \frac{1}{N} \sum_{i=1}^N m_i \\
\sigma &= \frac{1}{N-1} \sum_{i=1}^N (\mu - m_i)^2
\end{align}

where $m_i$ is the slope of the i-th linecut, and $N$ is the number of linecuts.
The distribution for the speed values can be seen in Figure \ref{Supp_Fig_Vel}.

\section{Discussion of position sensors/scan window drift}
\begin{figure}[h!]
    \centering
    \includegraphics[width=0.95\linewidth]{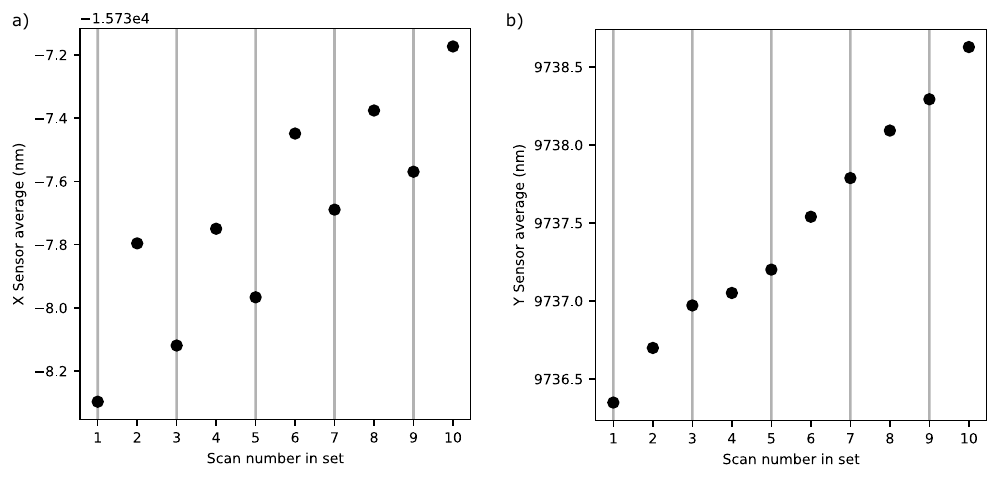}
    \caption{
        \label{Supp_Sensor_Avg}
        \textbf{Drifting position sensors/ scan window.} 
        The average value of (a) the X position sensor and (b) the Y position sensor. Points on the vertical lines denote scans with slow scan direction of positive slow axis); scans with slow scan direction of negative slow axis are interleaved. Here the slow axis is X. In (a), we see an alternating behavior in the average value of the X sensor. In (b), we see a constant increase in the average value of the Y sensor. We interpret these changes to be representative of a slow change in the center of the scan window.
        }
\end{figure}

Our analysis of drift assumes that the scanning probe always rasters the same window above the sample when performing successive scans. We found that the average values of the X/Y position sensors were slowly changing during successive scans. An example of this behavior is shown in Figure \ref{Supp_Sensor_Avg}. We interpret the average value of the X/Y position sensors as being representative of the center of the scan window. This means our scan window is not exactly the same between the successive scans. In order to compare equivalent scan windows for calibrating drift, we only analyze regions of the sample where there is overlap between all the scan windows.

\section{Explanation of drift velocity estimation}

\begin{figure}
    \centering
    \includegraphics{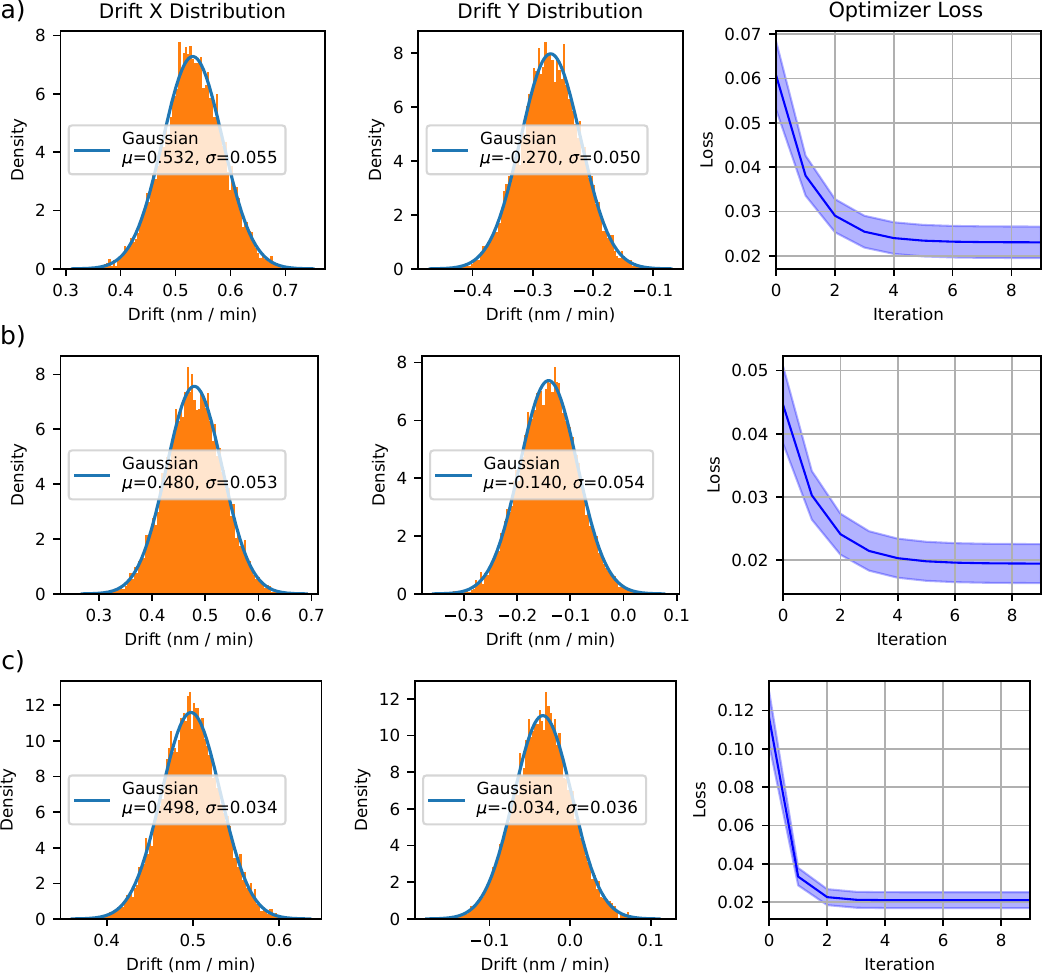}
    \caption{The distribution of the optimal drift values and the loss values for the optimizer after sampling the Peak position and tip velocities 10,000 times for the slow x axis scan.
    a) shows the Drift distribution for the slow axis Y, b) for slow axis X and c) for another slow axis X scan.
    The loss values for the optimizer are displayed to show that the parameters have converged and the optimal value has been found.
    The loss used is the Mean Squared Error (MSE).
    The loss did not converge to 0, but this is to be expected as the peak positions are noisy, thus it is not possible to reach an MSE loss of 0.
    We fitted gaussians to the distributions of the drift to extract the uncertainties for the drift.
    }
    \label{fig:drift_supp}
\end{figure}

The values for the drift can be estimated by solving the following formula:

\begin{align}
T^{-1}(V_{i_{fast}},V_{i_{slow}}, D_{fast}, D_{slow}) P_{j,i} = T^{-1}(V_{{i+1}_{fast}},V_{{i+1}_{slow}}, D_{fast}, D_{slow}) P_{j,i+1}\\
\forall i,j
\end{align}

where $T$ is the corresponding drift transformation matrix, $V_{i_{fast}}$ and $V_{i_{slow}}$ are the fast and slow tip velocities of the i-th scan, and $P_{j,i}$ is the j-th autocorrelation peak of the i-th scan.
Each scan has 6 autocorrelation peaks, which are the first 6 peaks measured from the middle.

These peaks have an associated uncertainty in their position of $\sigma = \pm \frac{1.5}{\sqrt{3}}$ pixels. This value is chosen because multiple sources of uncertainty influence the true uncertainty.
By choosing $\pm \frac{1.5}{\sqrt{3}}$ pixels, we effectively assert that the true position is uniformly distributed in a 3x3 pixel square around the detected peak position.
This is then converted to a Gaussian sigma value by dividing half the range of the uniform distribution by $\sqrt{3}$.

To solve for the drift velocity formula, it is rewritten into the following objective:

\begin{equation}
\text{minimize}\; \sum_{i,j} T_i^{-1} P_{i,j} - T_{i+1}^{-1}P_{j,i+1}
\label{optim_drift}
\end{equation}

This objective is optimized using gradient-based optimization.
This alone does not yield uncertainties on the parameters.
To propagate the uncertainties of the autocorrelation peaks locations and the uncertainties on the tip velocities to the drift, we use Monte Carlo simulation.
The idea behind Monte Carlo simulation is to treat the input parameters as random variables (RVs) and sample from them according to their distribution.
By sampling the RVs, we always get a different optimal value.
The more we sample, the more accurately we can determine the distribution of the optimal drift parameter, giving us a value for the uncertainties on the drift.
We sample the RVs 10,000 times and optimize the objective \ref{optim_drift}.
The resulting distribution of the optimal drift values can be seen in Figure \ref{fig:drift_supp}.

\section{Explanation of Twist Angle Estimation}

\begin{figure}
    \centering
    \includegraphics[width=\linewidth]{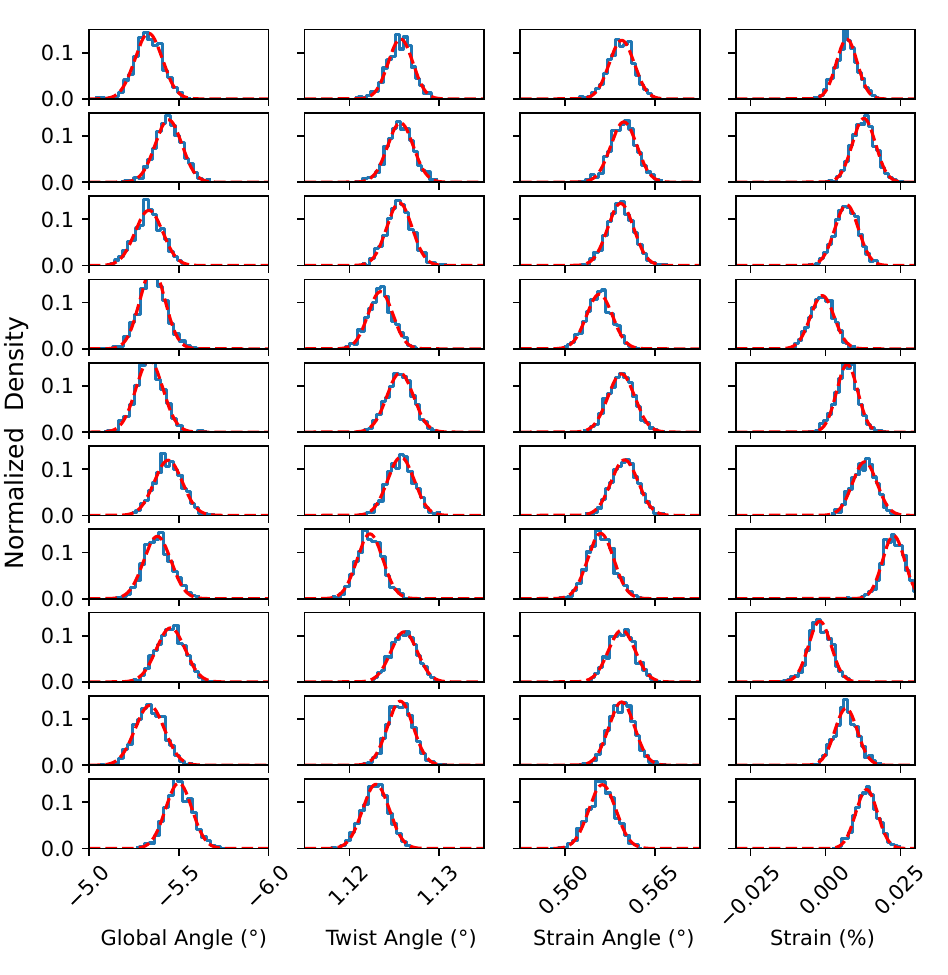}
    \caption{The distribution of the parameters for the global angle ($\alpha$), twist angle ($\theta$), strain angle ($\phi$) and strain ($\epsilon$) with plotted gaussian distributions for all scans with slow X axis.
    Each row of plots indicates one scan, while each column indicates the parameters.
    For each scan we ran the monte carlo simulation 1000 times to obtain statistics over the parameters.
    We can see that the final optimal parameters are gaussian distributed.
    }
    \label{fig:parameters}
\end{figure}

The global angle ($\alpha$), twist angle ($\theta$), strain angle ($\phi$) and heterostrain ($\epsilon$) can be obtained by solving the following formula:

\begin{align}
R(\theta) =
\begin{bmatrix}
\cos(\theta) & -\sin(\theta) \\
\sin(\theta) & \cos(\theta)
\end{bmatrix}
\end{align}

\begin{align}
S(\theta, \epsilon) =
R(\theta)^T
\left(
I +
\epsilon \cdot
\begin{bmatrix}
1 & 0 \\
0 & -0.16
\end{bmatrix}
\right)
R(\theta)
\end{align}

\begin{equation}
R(\alpha) [R(\theta)S(\phi, \epsilon) G_i - G_i ]= M_i \;\forall i\in[1,2]
\label{optim_1}
\end{equation}

where $G_i$ are the graphene lattice vectors in K-space, $M_i$ are the moiré vectors in K-space, $\alpha$ is the global rotation of the lattice vectors, $\theta$ is the relative twist angle of the lattices, $\phi$ is the strain angle, and $\epsilon$ is the strain.

As the position of the moir\'e peaks is noisy, we use not just 2 Moiré peaks from the autocorrelation but the 6 closest to the center. These peaks represent the moir\'e vectors in real space, so we first transform them to k-space using the following formula:

\begin{align}
A = \frac{1}{2\pi}B^{-1}
\end{align}

where $A$ is a matrix with the K-space vectors as columns and $B$ is a matrix with real space vectors as rows.

As it is only possible to transform pairs of moiré vectors to K-space, we use all possible independent pairs of Moiré vectors, totaling 12. This gives us a total of 48 equations to solve for. Equation \ref{optim_1} is then rewritten into the following objective:

\begin{align}
\text{minimize}\sum_i  R(\alpha) [R(\theta)S(\phi, \epsilon) - I ] G_i - M_i
\end{align}

This objective is optimized using a gradient-based optimizer.
We again use the Monte Carlo method to propagate the uncertainties to the parameters.
The resulting distribution for $\alpha$, $\theta$, $\phi$ and $\epsilon$ can be seen in figure \ref{fig:parameters}.

Monte Carlo generally works very well for non-degenerate problems, but due to very low strain ($\approx 0.01\%$), the strain matrix ($S(\phi,\epsilon)$) collapses to an identity matrix ($I$). This results in the following formula for the strain matrix:

\begin{align}
S(\theta) = R(\theta)^T R(\theta) = I
\end{align}

Here we see that any value for the strain angle ($\theta$) is a valid parameter, as any value will result in an identity matrix. In this case, the Monte Carlo simulation does not return a valid uncertainty for the strain angle.
Furthermore, this makes the optimized value dependent on the starting parameters, making the computed uncertainties for the strain angle unreliable at such low strain.
As the other parameters ($\alpha,\phi,\epsilon$) are not degenerate, their estimates of the uncertainties are reliable.

\clearpage
\section{Additional Data: varying line scan rate}
\begin{figure}[h!]
    \centering
    \includegraphics[width=\linewidth]{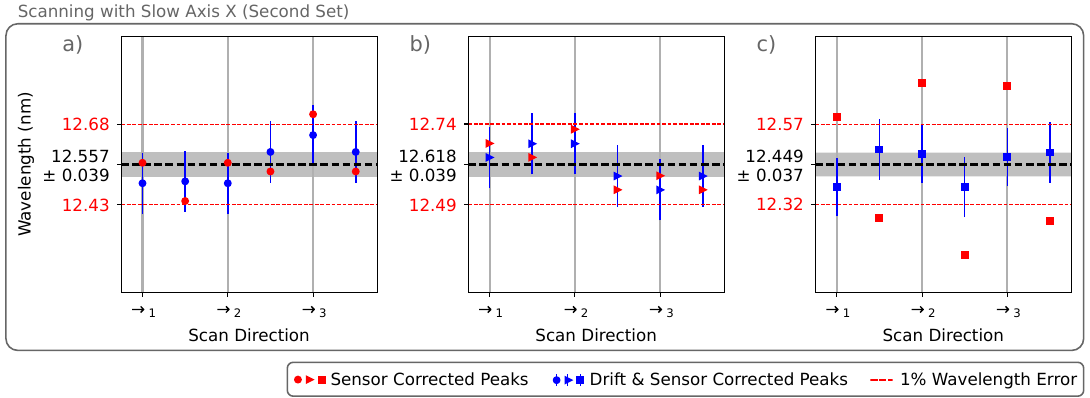}
    \caption{
    The extracted moir\'e wavelengths for measurement Set 3 of Table I in the main text. Data were acquired immediately after Set 2 with slow axis X and a line scan rate of 4 Hz.  Before drift correction (red markers), the extracted wavelengths for (c) are outside the 1\% wavelength error (compare to Figure 2f of the main text.) This is expected since individual scans in Set 3 took twice as long compared to Set 2, so the effect of relative drift between sample and probe is enhanced. After drift correction, the average wavelengths of moir\'e lattice vectors are are consistent between Sets 2 and 3.}
    \label{D3_Scan_Wavelengths}
\end{figure}

\begin{figure}[h!]
    \centering
    \includegraphics[width=\linewidth]{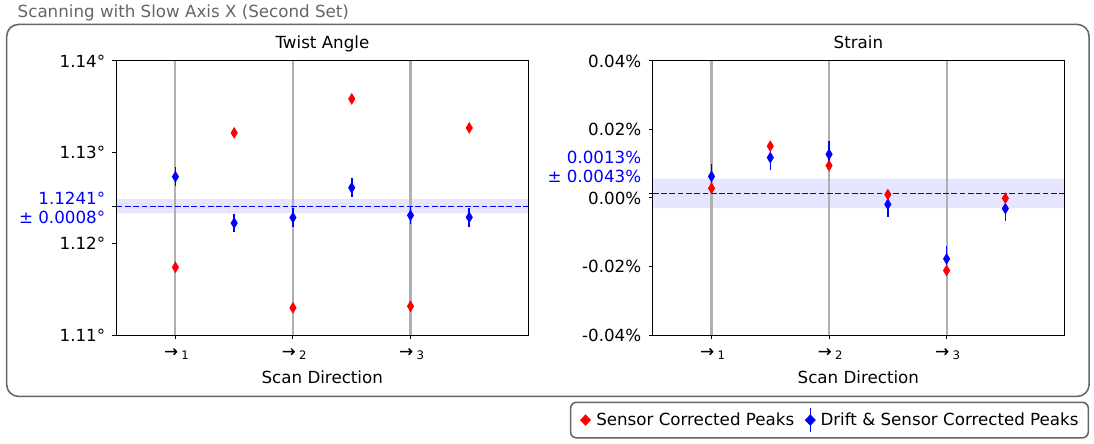}
    \caption{
    The extracted TBG twist angle and heterostrain magnitude for measurement Set 3 (Figure \ref{D3_Scan_Wavelengths}) before and after drift correction. The average TBG twist angle and heterostrain is consistent with Sets 1 and 2 after drift correction (see main text).}
    \label{D3_Scan_Moire_Structure}
\end{figure}
\end{document}